\documentstyle[prd,preprint,aps,psfig]{revtex}
\begin{document}
\draft
\title{TEMPERATURE EVOLUTION LAW OF IMPERFECT RELATIVISTIC FLUIDS}
\author{R. Silva \thanks{Electronic address: raimundo@dfte.ufrn.br} J. A. S. Lima\thanks{Electronic address: limajas@dfte.ufrn.br} and M. O. Calv\~ao\thanks{Electronic address: orca@if.ufrj.br}}
\address{Departamento de F\'{\i}sica Te\'orica e Experimental\\
Universidade Federal do Rio Grande do Norte, C. P. 1641\\
59072-970, Natal-RN, Brazil}
\address{~\\$^2$Instituto de F\'{\i}sica,
\\Universidade Federal do Rio de Janeiro,
\\21945-970, Rio de Janeiro, RJ, Brazil}

\maketitle
\begin{abstract}
The first-order general relativistic theory of a generic
dissipative (heat-conducting, viscous, particle-creating)
fluid is rediscussed from a unified covariant frame-independent
point of view. By generalizing some previous works in the literature, we derive a formula for the temperature variation rate, which is valid both in Eckart's (particle) and in the Landau-Lifshitz (energy) frames.  Particular attention is paid to the case of gravitational particle creation and its possible cross-effect with the bulk viscosity mechanism.
\end{abstract}


\section{Introduction}

The formulation of the cosmological problem, for any particular
model, is based on a set of phenomenological quantities such as energy density, pressure and temperature, which in principle, need to be defined with respect to local observers. In this regard, as pointed out by G\'eheniau
and collaborators\cite{Geheniau87}, the interdisciplinary field of
cosmology is the natural inheritor of the concepts and methods
ordinarily employed in hydrodynamics or, more generally, in
non-equilibrium thermodynamics\cite{deGroot84,Kreuzer81}.

Besides being essential to complete the aforementioned
phenomenological description, the thermodynamic analysis 
of cosmological models may also have interesting observational 
consequences. In principle, the dimensionless entropy content of the (observable) universe, a huge number of the order of ${10}^{87}$,
may have been generated by dissipative mechanisms like heat
conduction, viscosity, "chemical reactions", diffusion, photon 
creation, etc., taking place in an inhomogeneous and anisotropic
stage of the early universe \cite{Weinberg71,Weinberg72}. In this case,
after some few minutes, the gradients of the main physical quantities will be constrained from primordial nucleosynthesis studies. Later on, at the time of recombination, these irreversible mechanisms may also contribute to the temperature anisotropy of the cosmic background radiation. Indeed, although severely restricted from COBE measurements, the presence of the classical dissipative mechanism are naturally expected in a non-homogeneous quasi FRW universe, and, in principle, their specific signature must be somewhat different of the standard Sunyaev-Zeldovich and Sachs-Wolf effects\cite{PE}. 

In order to unify in a single and coherent scheme all irreversible
phenomena occurring in a simple fluid or in mixtures, standard
non-equilibrium thermodynamics works with two basic
ideas\cite{deGroot84,Kreuzer81}. The first one is the local
equilibrium hypothesis, whose mathematical expression is given by
the equilibrium Gibbs law in its local form. It implies that, out
of equilibrium,  the basic state functions such as the entropy, depend
locally on the same set of thermodynamic variables as in
equilibrium. In particular, the usual thermodynamic temperature
and pressure concepts are maintained in the non-equilibrium
regime. The second idea is that, in the presence of dissipative
processes, there is a local entropy source strength $\tau$
(entropy variation per unit volume and unit time), which, by the
second law of thermodynamics, is always non-negative.
Mathematically, it takes the form of a balance equation with
$\tau$ as a source term.

By combining such assumptions with the fluid equations of motion,
one finds an expression for the entropy source strength, as well
as for the constitutive (phenomenological) relations themselves.
The usual equilibrium theory is readily recovered by taking the
limit of vanishing entropy production rate.

This approach has systematically been applied at the classical
level\cite{deGroot84,Kreuzer81} as well as in the
special-relativistic domain\cite{Weinberg72}-\cite{Dixon78}. Its
extension to the general-relativistic framework is
straightforward, provided the gravitational field varies slowly
over the mean free path or the mean free time of the fluid
particles\cite{Weinberg71}. Formally, since, under these
conditions, the equivalence principle should hold, the generally
covariant equations may be established by the usual minimal
coupling recipe, that is, the replacement of usual derivatives by
covariant derivatives and the replacement of the Minkowski metric
$\eta_{{\alpha}{\beta}}$ by its Riemannian counterpart
$g_{{\alpha}{\beta}}$. However, as is widely known, in the
relativistic case, there is an ambiguity related to the possible
choices of the macroscopic hydrodynamic four-velocity. In Eckart's
formulation\cite{Eckart40}, the four-velocity is directly related
to the particle flux, while, in Landau-Lifshitz's
approach\cite{Landau59}, it is directly related to the energy
flux. In principle, a general treatment should be able to deal with any of these ``gauge'' choices. 

In this paper, we consider a unified covariant description of a heat-conducting viscous simple fluid with particle creation. The theories of Eckart and Landau-Lifshitz will be seen to be special cases of this general formulation. We also derive a new temperature evolution law which holds for any particular choice of the macroscopic hydrodynamic four-velocity.

\hspace{.3in}

\section{General Relativistic Theory of Fluids}

The thermodynamic state of a relativistic simple fluid is
characterized by an energy-momentum tensor $T^{{\alpha}{\beta}}$,
a particle current $N^{\alpha}$ and an entropy current
$S^{\alpha}$. The fundamental equations of motion are expressed by
the conservation law (semi-colon denotes covariant derivative) of
energy-momentum and the equation of balance for the particle
number\cite{Calvao92}

\begin{equation} \label{eq:1}
{T^{{\alpha\beta}}}_{;\beta} = 0 \quad ,
\end{equation}
\begin{equation} \label{eq:2}
{N^{{\alpha}}}_{;\alpha} = \Psi \quad,
\end{equation}

\noindent where $\Psi$ is a particle source ($\Psi>0$) or sink
($\Psi<0$) term. The second law of thermodynamics requires that
the entropy source strength be non-negative

\begin{equation} \label{entropyflux}
{S^{{\alpha}}}_{;\alpha}=\tau\geq 0 \quad,
\end{equation}

\noindent where $\tau=0$ describes a non-dissipative state, and
$\tau>0$ denotes a dissipative state. A perfect fluid always
evolves through non-dissipative (equilibrium) states, whereas an
imperfect fluid typically evolves through dissipative
(non-equilibrium) states (see, however, \cite{Bedran93}).

\subsection {Adiabatic Limit}

By choosing an arbitrary hydrodynamic frame of reference, whose
four-velocity obeys $u^\alpha u_\alpha = 1,$ the primary variables
$T^{\alpha\beta}, N^\alpha$ and $S^\alpha$ take the following
forms \cite{Weinberg72,Dixon78}

\begin {equation}\label{energy}
T^{\alpha \beta} = \rho u^{\alpha} u^{\beta} - ph^{\alpha \beta},
\end{equation}
\begin{equation} \label{particle}
N^{\alpha} = nu^{\alpha},
\end{equation}
\begin {equation} \label{entropy}
S^{\alpha} = n\sigma u^{\alpha},
\end{equation}

\noindent where the tensor $h^{\alpha \beta} :=
u^{\alpha}u^{\beta} - g^{\alpha \beta}$ is the usual projector
onto the local rest space of $u^\alpha$. The variables $\rho,$
$p$, $n$ and $\sigma$ stand respectively for the energy density,
thermostatic pressure, particle number density and specific
entropy (per particle), and are related by the so-called Gibbs law
\cite{Weinberg72,Dixon78}

\begin{equation} \label{gibbs} nTd\sigma = d\rho - {\rho + p \over
n}dn \quad. \end{equation}

As a particular result of the general theory,
(\ref{energy})-(\ref{entropy}) imply that the entropy source
strength defined by (\ref{entropyflux}) vanishes identically,
\begin{equation}\label{equilibrium}
{S^{\alpha}}_{;\alpha}=\tau=0\quad,
\end{equation}

\noindent as should be expected for thermal equilibrium states.

\subsection{Non-Equilibrium States}

In principle, the inclusion of dissipative processes such as heat
conduction and viscosity, requires additional terms in the primary
variables describing a perfect fluid. However, unlike in the
adiabatic limit case, the presence of a heat transfer poses a
problem regarding the definition of the hydrodynamic four-velocity
$u^{\alpha}$. It is necessary to specify whether $u^{\alpha}$ is
the four-velocity of the energy transport or particle transport.
In Eckart's formulation, $u^{\alpha}$ is identified with the
four-velocity of particle transport (particle
frame)\cite{Eckart40}. In the approach of Landau-Lifshitz,
$u^{\alpha}$ is defined as the four-velocity of energy transport
(energy frame)\cite{Landau59}. Formally, the particle frame is the
unique unit time-like vector parallel to $N^\alpha$, whereas the
energy frame is the unique unit time-like eigenvector of
$T^{\alpha\beta}$. Both theories assume that, for weak space-time
gradients, the basic quantities contain no terms higher than first
order in deviations from equilibrium.

In the presence of irreversible processes, we must add small terms
$\Delta T^{\alpha\beta}$ and $\Delta N^\alpha$ in (\ref{energy})
and (\ref{particle}), which are restricted by the second law of
thermodynamics (\ref{entropyflux})
\begin{equation} \label{energypertub}
T^{\alpha \beta} = \rho u^{\alpha} u^{\beta} - p h^{\alpha \beta}
+ \Delta T^{\alpha\beta}\quad,
\end{equation}
\begin{equation} \label{eq:10}
N^\alpha=n u^\alpha + \Delta N^\alpha\quad,
\end{equation}

\noindent Usually, at this point, one specifies whether the Eckart
or Landau-Lifshitz approach  will be adopted. However, this is not
necessary in the covariant frame-independent formulation presented
here since these theories will be recovered as particular cases.

As before the fluid motion equations are contained in (\ref{eq:1})
and (\ref{eq:2}), which express the energy conservation law and
the balance equation for the particle number, respectively.
Therefore, differentiating the expression (\ref{energypertub}) and
projecting it on the direction of the four-velocity $u_\alpha$,
one finds

\begin{equation} \label{eq:12}
{u_\alpha T^{\alpha\beta}}_{; \beta} = \dot{\rho}+(\rho+p)\theta+
{u_{\alpha}\Delta T^{\alpha\beta}}_{;\beta}=0\quad,
\end{equation}
and from (\ref{eq:2})
\begin{equation}\label{balpart}
{N^{\alpha}}_{;\alpha}=\dot{n}+n\theta + {\Delta
N^{\alpha}}_{;\alpha}=\Psi\quad,
\end{equation}
where an overdot means the derivative along the world lines of the
fluid volume element, e.g., $\dot{\rho}=u^{\beta}\rho_{;\beta}$,
and $\theta={u^{\beta}}_{;\beta}$ is the expansion rate of the
fluid.

Now, taking the covariant derivative of (\ref{gibbs}) along the
world lines of the fluid volume element, and making use of
equations (\ref{eq:12}) and (\ref{balpart}), we obtain

\begin{equation}\label{eq:14}
T(n\sigma u^{\beta})_{;\beta}=-u^{\alpha}\Delta
{T^{\alpha\beta}}_{;\beta}- \mu\Psi +\mu\Delta
{N^\beta}_{;\beta}\quad,
\end{equation}
where $\mu$ is chemical potencial defined by Euler's relation
\begin{equation} \label{eq:17}
\mu={\rho+p\over n}-\sigma T\quad.
\end{equation}
Defining the entropy flux
\begin{equation} \label{eq:21}
S^\beta =n \sigma u^\beta - {\mu\over T}\Delta N^{\beta}+
{u_{\alpha}\over T}\Delta T^{\alpha\beta}\quad,
\end{equation}
we see from (\ref{eq:14}) that the entropy source strength assumes
the following form
\begin{equation}\label{4}
{S^{\beta}}_{;\beta}=\left({u_{\alpha;\beta}\over T}-
{T_{;\beta}u_\alpha \over T^2}\right)\Delta T^{\alpha\beta}-
\left({\mu_{;\beta}\over T}-{\mu T_{;\beta}\over T^2}\right)
\Delta N^{\beta}-{\mu\Psi\over T}\quad,
\end{equation}
which is a function dependent only on the dissipative fluxes.

At the level of the primary fluxes, the effect of the dissipative
processes is to add up new fields for the energy-momentum tensor
and the particle flux
vector\cite{Eckart40,Weinberg72,Calvao92,Hiscok82}, which must be
constrained by the second law. We have
\begin{equation}\label{2}
\Delta T^{\alpha\beta}=-({\Pi}+p_c)h^{\alpha\beta}+q^\alpha
u^\beta+ q^\beta u^\alpha + \Pi^{\alpha\beta}\quad,
\end{equation}
and
\begin{equation}\label{3}
\Delta N^{\beta}=\nu^{\beta}\quad,
\end{equation}
where the five additional fields $\Pi,p_c,q^{\alpha},\nu^{\beta}$
and $\Pi^{\alpha\beta}$ are, respectively, the bulk viscous
pressure, the creation pressure (due to the gravitational matter
creation\cite{Calvao92}), the heat flow, the particle drift, and
the shear viscosity stress. These fields describe the deviations
from equilibrium within the fluid and satisfy the following
constraints
\begin{equation}\label{5}
u^\alpha q_\alpha =u^\alpha\nu_{\alpha}= u^\alpha \Pi_{\alpha
\beta} = g_{\alpha \beta} \Pi^{\alpha \beta} = \Pi^{[\alpha
\beta]} = 0 \quad,
\end{equation}

\noindent where the square brackets denote anti-symmetrization and
round brackets, below, symmetrization.

As one may check, inserting equations (\ref{2}) and (\ref{3}) into
(\ref{4}), and using (\ref{5}), it is readily seen that
\begin{equation} \label{eq:22}
{S^\alpha}_{;\alpha}= -{\Pi\theta\over T} - {p_c\theta\over T} -
{\mu\Psi\over T} - \left({T_{;\beta}\over T^2} -
{\dot{u}_{\beta}\over T}\right)q^{\beta}-\left({\mu_{;\beta}\over
T}- {\mu T_{;\beta}\over T^2} \right)\nu^\beta+{u_{\alpha;\beta}
\Pi^{\alpha\beta}\over T}
\end{equation}
On the other hand, the covariant derivative of $u^\alpha$ may be
decomposed as \cite{Ellis73}
\begin{equation} \label{eq:27}
u_{\alpha;\beta} = \sigma_{\alpha\beta} + {1\over 3} \theta
h_{\alpha\beta} + \dot u_{\alpha} u_\beta + \omega_{\alpha\beta}
\quad,
\end{equation}
where
\begin{equation} \label{eq:28}
\sigma_{\alpha\beta} = {1\over 2} h^\mu _\alpha h^\nu _\beta
(u_{(\mu;\nu)}- {2\over 3} \theta h_{\mu\nu}) \quad,
\end{equation}
and
\begin{equation} \label{eq:29}
\omega_{\alpha\beta} = {1\over 2} h^\mu _\alpha h^\nu _\beta
u_{[\nu,\mu]} \quad,
\end{equation}
are respectively the symmetric traceless shear tensor and the
vorticity tensor. Inserting (\ref{eq:27}) into equation
(\ref{eq:22}), we may rewrite the entropy source as
\begin{equation}\label{18}
{S^{\alpha}}_{;\alpha} =
\tau_{s}-{q_{\alpha}h^{\alpha\beta}(T_{;\beta}-
T\dot{u}_{\beta})\over T^2}-\left({\mu_{;\beta}\over T}- {\mu
T_{;\beta}\over T^2} \right)\nu^\beta+
{\Pi^{\alpha\beta}\sigma_{\alpha\beta}\over T}\quad,
\end{equation}
where $\tau_{s}$ describes the entropy source strength due to the
scalar irreversible processes
\begin{equation}
\tau_{s}= -{\Pi\theta\over T} - {p_c\theta\over T}- {\mu\Psi\over
T}\quad.
\end{equation}
Hence, to the first-order of approximation, (\ref{18}) will be
consistent with the second law of thermodynamics if the
phenomenological relations among the vector and tensor dissipative
fluxes and thermodynamic forces are taken to be

\begin{equation} \label{eq:32a}
q^\alpha = \chi\phi^\alpha\quad;\quad\phi^\alpha = h^{\alpha\beta}
(T_{;\beta} - T\dot u_\beta)\quad,
\end{equation}
\begin{equation} \label{eq:34}
\Pi^{\alpha\beta} = \eta\sigma^{\alpha\beta}\quad,
\end{equation}
\begin{equation} \label{eq:35}
\nu^\alpha=\zeta \lambda^\alpha;\quad \lambda^\alpha=
h^{\alpha\beta}\left({\mu\over T}\right)_{;\beta}\quad,
\end{equation}
where $\chi, \eta,$ and $\zeta$ stand, respectively, for thermal
conductivity, shear viscosity and ``difusion" coefficients. Using
(\ref{eq:32a})-(\ref{eq:35}), the entropy source strength
(\ref{18}) becomes

\begin{equation} \label{eq:36}
{S^\alpha}_{;\alpha} = \tau_{s}-
{{\chi}\phi_{\alpha}{\phi^{\alpha}} \over T^2} -
{\zeta\lambda_\alpha\lambda^\alpha\over T^2} +
{\eta\sigma_{\alpha\beta}\sigma^{\alpha\beta}\over T}\quad,
\end{equation}
and since the heat flow and the particle drift are space-like
vectors $(\phi^\alpha \phi_\alpha < 0,
\lambda^\alpha\lambda_\alpha < 0)$, the second law of
thermodynamics will be satisfied if, for any configuration of the
fluid, $\chi$, $\eta$ and $\zeta$ are positive. In order to treat
the scalar irreversible processes, we observe that they are fluxes
with same tensor rank (scalars), and in the first-order of
approximation this give rise to cross effects
\cite{deGroot84,Gariel95}. In order to describe this cross effect,
we propose the following relations among fluxes and thermodynamic
forces,
\begin{equation}\label{flu1}
\Pi= -\xi_{11}\left({\Psi\over\theta}\right) - \xi_{12}\theta
\quad,
\end{equation}
and
\begin{equation}\label{flu2}
p_c= -\xi_{21}\theta - \xi_{22}\left({\Psi\over\theta}\right)
\quad,
\end{equation}
where $\Pi$, $p_c$ are fluxes and
$\left({\Psi\over\theta}\right)$, $\theta$ are thermodynamic
forces, with $\xi_{ij}>0$ ($i,j=1,2$) phenomenological
coefficients. When the particle source strength $\Psi$ is zero, we
see from (\ref{flu1})-(\ref{flu2}) that $\Pi=-\xi_{12}\theta$ and
$p_c=-\xi_{21}\theta$. Recalling that the phenomenological
coefficients are supposed to obey Onsager's reciprocity relations,
it follows that $\xi_{12}=\xi_{21}$. This means that in this
particular case (no particle creation) the creation pressure and
bulk viscosity reduce to the same process. In the general case,
the entropy source (\ref{eq:36}) can be written as
\begin{equation}\label{entrfinal}
{S^\alpha}_{;\alpha} = - {{\chi}\phi_{\alpha}{\phi^{\alpha}} \over
T^2} + {\eta\sigma_{\alpha\beta}\sigma^{\alpha\beta}\over T}-
{\zeta\lambda_\alpha\lambda^\alpha\over\
T^2}+{2\xi_{12}\theta^2\over T} +{(\xi_{11} + \xi_{22}-{\mu})\Psi
\over T}
\end{equation}
We see that the second law of thermodynamics will be satisfied for
any configurations of the fluid if
$\chi$,$\eta$,$\zeta$,$\xi_{12}$ are positive definite. We also
see that the remaining coefficients $\xi_{11}$ and $\xi_{22}$ must
satisfy the following inequalities: $\xi_{11}$ + $\xi_{22} > \mu$
if $\Psi>0$, and $\xi_{11}$ + $\xi_{22} < \mu$ if $\Psi<0$ (cf.\
\cite{Calvao92}).

Concerning the results presented in the reference \cite{Gariel95},
we have the following remarks. First, we observe that the flux
associated with the particle creation rate, namely, the creation
pressure was not considered. Second, the authors have used $\Pi$
as a force and not as, we think, a thermodynamic flux. Another
question is related to the hypothesis implicitly adopted in their
paper, that the creation pressure and the bulk viscous pressure
are the same ($\Pi=p_c$). In particular, from the phenomenological
laws obtained there, we see that for photons ($\mu=0$), the
creation pressure is given by $p_c={\xi_{12}\over\xi_{22}}\Psi$;
however for the static fluid $\theta=0$, and using the energy
conservation, we find that $\dot{\rho}=0$. These results seem to
be physically inconsistent, because we have energy density $\rho$
constant and simultaneously  photon creation, which should be
responsible for a variation of $\rho$. Note also that, in the
adiabatic limit, the entropy source strength vanishes, as expected
(see \ref{equilibrium}).

Summing up, the basic set of equations governing the first-order
theory of dissipative simple fluids in a general formulation are:
the energy conservation law (\ref{eq:1}) and equation of balance
(\ref{eq:2}) with $T^{\alpha\beta}$ and $N^\alpha$ given by
(\ref{energypertub}) and (\ref{eq:10});  the Gibbs law
(\ref{gibbs}), the constitutive equations
(\ref{eq:32a})-(\ref{eq:35}) and the constitutive equation for a
scalar process (\ref{flu1})-(\ref{flu2}) with the fluxes
constrained by (\ref{5}) and linked with the entropy source
strength by (\ref{entrfinal}). Notice that such a system is
underdetermined since, in principle, there are fifteen independent
equations and seventeen unknowns; more specifically, as in the
classical theory, the time behavior of the relativistic
one-component fluid can only be determined, for specified initial
boundary conditions, by adding two equations of state, say

\begin{equation} \label{eq:37}
p = p(n, T)\quad,
\end{equation}
and
\begin{equation} \label{eq:38}
\rho = \rho(n, T)\quad.
\end{equation}
We notice that, although sucessful in revealing the physics
underlying a large class of phenomena, the first-order theories
present some experimental and theoretical drawbacks. In its
classical version, the linear constitutive equations
(\ref{eq:32a})-(\ref{eq:35}) and (\ref{flu1})-(\ref{flu2}) are not
adequate at high frequencies or short wave lengths as manifested
in experiments on ultrasound propagation in rarefied gases and on
neutron scattering in liquids \cite{Jou89}. Besides they also
allow the propagation of perturbations with arbitrarily high
speeds, which, although perhaps merely unsatisfactory on classical
grounds, is completely unacceptable from a relativistic point of
view; furthermore, they do not have a well-posed Cauchy problem
and their equilibrium states are not stable. Several authors have
formulated relativistic second-order theories which circumvent
these defficiencies\cite{Dixon78,Israel76,Pavon82,Hiscock83}. In a
forthcoming paper\cite{Silva00}, we intend to extend our
considerations to this class of theories.

\section {Temperature Evolution Law}

Let us now derive the general equation for the temperature law in
a relativistic fluid taking into account all the irreversible
processes. In what follows we consider that the energy density and
the equilibium pressure are functions of the thermodynamic
variables $n$ and $T$. Differentiating $\rho = \rho(n,T)$ along
the world lines of the fluid volume element one finds
\begin{equation}\label{rho.}
\dot\rho= \left({\partial\rho \over\partial T}\right)_n\dot n +
\left({\partial\rho \over\partial n}\right)_T\dot T \quad,
\end{equation}
and combining with the energy conservation law (\ref{eq:12}) and
the balance equation (\ref{balpart}), we obtain
\begin{equation}\label{eq:47}
({\partial \rho\over\partial T})_n \dot T =\left[n({\partial
\rho\over\partial n})_T -\rho-p\right]\theta-({\partial
\rho\over\partial n})_T(\Psi-\Delta {N^\alpha}_{;\alpha})-
u_\alpha\Delta {T^{\alpha\beta}}_{;\beta}\quad.
\end{equation}
Since $d\sigma$ must be an exact differential, (\ref{gibbs}) leads
to the thermodynamic relation
\begin{equation} \label{eq:480}
\left[{\partial\over\partial T} \left( {1\over
nT}\left[\left({\partial\rho\over\partial n}\right)_T-
\left({\rho+p\over n}\right)\right] \right) \right]_n=
\left[{\partial\over\partial n} \left( {1\over
nT}\left({\partial\rho\over\partial T}\right)_n
\right)\right]_T\quad,
\end{equation}
or equivalently,
\begin{equation}\label{eq:500}
T\left({\partial p\over\partial T}\right)_n=
\rho+p-n\left({\partial\rho\over\partial n}\right)_T\quad,
\end{equation}
and combining (\ref{eq:500}) with (\ref{eq:47}), we obtain the
general equation governing the variation of temperature
\begin{equation}\label{tempgeral}
{\dot T\over T} = -{\left({\partial p\over \partial
\rho}\right)_n{\theta}} - {1\over {T\left({\partial
\rho\over\partial T}\right)_n}} \left[\left({\partial \rho\over
\partial n}\right)_T(\Psi-\Delta {N^\alpha}_{;\alpha}) + u_{\alpha}
\Delta {T^{\alpha\beta}}_{;\beta}\right]\quad,
\end{equation}
where all dissipatives fluxes are described by $\Delta N^\alpha$
and $\Delta T^{\alpha\beta}$. In the adiabatic limit, the above
equation reduces to the temperature variation rate of a perfect
fluid
\begin{equation}\label{fluperf}
{\dot T\over T} = -{\left({\partial p\over \partial
\rho}\right)_n{\theta}}\quad.
\end{equation}

Consider now in (\ref{tempgeral}) the Landau-Lifshitz or energy
frame. In this case, the comoving observers do not see the
irreversible contribution of the energy flux, that is,
$q^\alpha=0$ in the energy-momentum tensor. As we have seen, there
is an additional contribution $\nu^{\alpha}$ in the particle flux
$\Delta N^\alpha$. We obtain
\begin{equation}  \label{eq:51}
{\dot T\over T} = -{({\partial p\over \partial
\rho})_n{\theta_{LL}} + {1\over {T{(\partial \rho\over\partial
T})_n}} \left[({\partial \rho\over
\partial n})_T (\nu^{\alpha}_{;\alpha}-\Psi) - (\Pi+p_c)\theta_{LL}+
\sigma_{\alpha\beta} \Pi^{\alpha\beta}\right]}
\end{equation}
where
\begin{equation}\label{a}
\theta_{LL}={\Psi-\dot{n}-{\nu^\alpha}_{;\alpha}\over n}
\end{equation}
is the expansion of the fluid in the Landau-Lifshitz frame.

Now let us consider the Eckart or particle frame formulation. In
this case, the comoving observers do not see the irreversible
contribution of particle drift, that is  $\Delta
N^\alpha=\nu^\alpha=0$ in the particle flux vector. We have
\begin{equation}  \label{eq:52}
{\dot T\over T} = -{({\partial p\over \partial
\rho})_n{\theta_{E}} + {1\over {T({\partial \rho\over\partial
T})_n}} \left[-(p_c+\Pi)\theta_{E} + \dot u_{\alpha}q^\alpha +
\sigma_{\alpha\beta}\Pi^{\alpha\beta}-
{q^{\beta}}_{;\beta}-({\partial\rho\over\partial n})_T
\Psi\right]}
\end{equation}
where
\begin{equation}\label{b}
\theta_E ={\Psi-\dot{n}\over n}
\end{equation}
is the expansion of the fluid in the Eckart frame.

It should be noticed that in the adiabatic limit, when all
dissipative fluxes are absent, equations (\ref{eq:51}) and
(\ref{eq:52}) reproduce the same temperature variation rate of a
perfect fluid. The same happens when the dissipative fluxes reduce to a creation pressure plus the bulk viscosity mechanism. For this homogeneous and isotropic dissipative simple fluid, the temperature law has previously been derived by Lima and Germano\cite{LG92} (see also Zimdahl\cite{ZIM97} to the case of a two-fluid mixture). We also emphasize that, for each formulation, the complete temperature evolution equation must play an important role in problems of astrophysics and cosmological interest envolving the classical non-equilibrium mechanisms. In the cosmological domain, for instance, a natural application of equations (\ref{eq:51}) and
(\ref{eq:52}), is related to the contributions of the irreversible processes to the anisotropy temperature of the cosmic background radiation during the decoupling time.

\end{document}